# PROMOTING INDUSTRY-UNIVERSITY PARTNERSHIP IN INFORMATION TECHNOLOGY

Anthony Afuwoqi [1], Hongyou Wu [2]

1. Department of Computer Science, Obafemi Awolowo University, NIGERIA

    Email: anthony@oauife.edu.ng
2. Department of Mathematics, Xi'an University of Arts and Science, Xi'an, CHINA

    Email: hongyou.wu@stu.xawl.edu.cn

**ABSTRACT**

It is becoming increasingly difficult for Nigerian universities to go it alone in terms of serving as a citadel of learning, coping with the huge wage bill and competing with their peers in other parts of the world, due to competitive, economic and other pressures. As a consequence, Nigerian universities are left with no option than to carry their industrial partners along in terms of research and development through the formation of partnerships for their mutual benefit. Since the industries are established for profit making and the universities for knowledge enhancement, such partnerships would help in spreading the costs in terms of provision of knowledge and costs of research. This paper discusses the various types of partnerships involving industries and universities, the benefits derived and a possible model for the working of such a partnership which could be adapted to other sectors and countries in sub-Saharan Africa.

**Introduction**

University research centres can be beneficial to industrial firms by providing them with a number of relationship alternatives that facilitate the advancement of knowledge and new technologies (Santoro and Chakrabarti, 2001), and also the foundation of scientific and technological literacy that is required in Research and Development, manufacturing, and support functions (Cannon and Sandler, 2000). To create the business leaders of the future, universities need access to significant resources that enable them to impart a quality education. These resources include personnel (faculty and support); infrastructure and equipment; training and professional development; industry advisors and other technology-based needs that are expensive to acquire (Kumar et al, 2002). In this dispensation of lowered funding for the educational sector in Nigeria which had consistently remained below the UNESCO approved minimum, universities are increasingly faced with budget constraints, late release of allocations from the National Universities Commission, increasing staffs overhead which all limit their abilities to acquire these resources.

The four main strands of activity that universities engage in where there is significant potential for developing partnerships with Corporate Nigeria (as the bigger industries are known in Nigeria ) are :-

- Teaching and Learning
- Research and Development
- Innovation and Knowledge Transfer
- Training and Retraining of students on Information Technology (IT) placements

The simplest way to understand the concept of partnership is to think of it as working together for mutual benefit (Goetsch and Davis, 2006) such as adding diversity, bringing in new ideas, concepts, technologies and market access (Polizzotto, 2000). They also result in reduced spending by the individual partners compared to if they did the task on their own, as well as reduced risk of failure.

Universities in Nigeria are also businesses in the service of educating and training people, while increasing the Nation's base of intellectual property through research efforts. Just as corporations in Nigeria today are under serious pressure to shore up their capital base, attract quality investments from within and outside the country and retain qualitative manpower, Nigerian Universities are today facing more pressures in terms of attracting more quality students out of the hundreds of thousands writing JAMB (The Joint Admissions and Matriculations Board, the body charged with conducting placement tests into Nigerian universities), produce more research results in the midst of scarce funds and ill-motivated lecturers, improve on their endowments, increase their internally-generated revenue base and even be more competitive in sporting competitions! Why not do what the big Corporate Nigeria have done; namely to form partnerships that will help mitigate some of the pressures they face?

The benefits students receive from these industry-academia programs [partnerships] go far beyond those that emerge from classroom lectures and textbook case studies. Instructors or lecturers who cite current, relevant research results are able to provide undergraduates, graduate students, and executives with direct examples of theoretical concepts applied to the real world (Stank, 2004). From the university's standpoint, the gains generated by collaborating with business are also significant. In concrete terms, universities are likely to receive one or more of the following from corporate partners: funds, materials, expertise, internships, and exposure to real-world business problems (Keefe, 2000). The benefits [to industry] of using university expertise for analytical assistance are considerable, particularly if equipment costs are prohibitive or if the analysis is performed irregularly. Academic institutions often provide a means for companies to validate and/or test new products and processes. This is what most companies think of when they hear the term "partnering" (Wright and Harris, 2006).

This work identifies the areas where universities and industries can work together to improve on Information Technology in the country and the attendant benefits to the collaborating partners. A model for such University-Industry partnership was equally proposed.

## 1. TYPES OF INDUSTRY-UNIVERSITY PARTNERSHIPS

An alliance or partnership is a relationship that is strategic or tactical, and that is entered into for mutual benefit by two or more parties having complementary business interests and goals (Segilnd, 1996). There are long established and natural links between universities and industries, not least because universities produce a pool of well-educated graduates and postgraduates from which the professional workforce is recruited. With the people come the ideas, skills and knowledge from which many companies derive their competitive edge.

The following are assets belonging to both parties in an industry-university partnership:

- Physical Assets/ Resources: laboratories, equipments and facilities
- Human Resources: highly skilled and experienced staff
- Other Knowledge Resources: information, database, libraries, processes, ideas, contacts, etc.
- Financial Resources: own research funds or access to public funds.

The differences between universities and industry lie in the extent and diversity of these resources and this is where collaboration is mutually beneficial. For example, companies may have considerably greater financial resource and many will have state-of-the-art IT facilities and resources, but few companies can match universities in terms of the breadth of their human and other knowledge resources. Even where the partners are evenly matched, the diversity of experience that can be drawn upon in a partnership and synergy that results, may give the competitive edge that leads to successful innovation.

There are many ways in which partnerships between universities and industry can be structured. CBI (2001) and Polizzotto (2000) identified the following arrangements:

**1.1 Contract Research:-** this is where the university conducts research in an IT-related area where they have previous expertise but the effort is focused, and funded, by the industrial partner. Usually a contract research team will comprise project-specific research and technical staff working under academic supervision with the industrial partner allowed to set the agenda for the project. The industry will generally wish to own all the results of the project that it is commissioning and to have the exclusive right to ues and exploit the results commercially.

**1.2 Collaborative Research:-** a partnership where the IT research goals are defined by all the partners and where all the partners (industrial and university) make an active contribution to the research activity is described as collaborative research. Here, all the project partners contribute to the costs of the research either in cash or in kind (although the universities will generally expect the industrial partners to cover the full cost of the project).

**1.3 Sponsored Research:-** this is where researchers put together proposals, which industry may then consider funding. The industrial partner does not necessarily feed creatively into the research, but is expected to fund the project. The results from the research efforts are shared and the academic or university partner is expected to publish them as early as possible. The university and the industrial partner negotiate details relating to publication and access rights to the research results such that it meets the requirements of both partners.

**1.4 Graduate Fellowships/ Studentships:-** sponsorships of a studentship, whether in full or in part, in an IT-related research topic, offers companies a means of gaining a foothold in an emerging area of computing. These studentship may include M.Sc./M.Tech., M.Phil or PhD. Although industries may help shape the student's research goals, those goals and the research methodology employed, must meet the university's requirements for the award of a degree and should reflect the student's endeavours.

**1.5 Student Projects and Placements:**- placement schemes can initiate a relationship between a university and a company. These placements could be in form of the Students Industrial Work Experience Scheme (SIWES) used in Nigerian universities as currently coordinated by the Industrial Training Fund (ITF). Students from the university spend some time in the company on gainful activity and benefit in terms of getting first-hand experience on real-life issues affecting the industries and even enhancing their employability.

**1.6 Sponsored and Honorary Posts and Secondments:-** in the same way that students can diffuse knowledge into companies, business people can do the same into universities. They can provide business/industry experience for teaching and innovation support and sometimes, applied research experience. While providing these benefits they can keep a watching brief on academic advance in computing and forge closer ties as precursors to future partnerships. Universities sometimes can offer honorary posts or secondments to specific industrial researchers which may be full-time or part-time and usually involve supervision or participation in a research project, teaching or occasional lectures. Academic secondments into industry are possible with the academic gaining an insight into and training in the latest technologies and methodologies used by the industry.

**1.7 University Consultancy and Associated Commercial Services:-** here, universities academic staff are encouraged to devote a proportion of their time to external work, such as providing consultancy services for external customers. Consultancy differs from research in that it does not have as its prime purpose the generation of new knowledge for the institution and is always on a short-term basis. Industry can sometimes pay to use spare capacity of the university's computing facilities and e-learning centres and sometimes, universities could purchase some specialized computing equipments for use by industrial partners at a fee.

**1.8 Clubs and Networks:**- clubs and networks of companies are also a form of partnership. They are often set up by an individual university or by an outside agency such as the Nigerian Computing Society, Shell Petroleum Development, First Bank Plc, etc. They could be national or international in character. Networks taking the form of research clubs usually have formal participation rules to ensure that benefits are distributed evenly and are only available to members. They give members an opportunity to keep abreast of an emerging area of computing and may also allow them collectively to help shape the direction of the research carried out by academics.

**1.9 Jobs:-** when particular companies and universities develop close relations, they get to know better the other's needs and what they have to offer, and have a good understanding of what they are getting from the other. This makes it possible for IT graduates to secure jobs with ease in such companies.

## 2. FRAMEWORK FOR DEVELOPING A UNIVERSITY-INDUSTRY PARTNERSHIP

To create a partnership, the partners need to have a shared vision on IT-related ideas or concepts which should address the specific needs of all partners in the partnership. The nature of industry-university partnerships is not new and several models or frameworks for such mutual cooperation have been proposed and acted upon over the years. Zachman (1987) for instance proposed a framework which was subsequently adapted by Whitten et al (2001) as the foundation for developing the business partnership framework. The major responsibility of IT/universities is to provide trained manpower to meet the needs of industry/ professional sector. Their output must cater to the actual need, requirement and expectation of the local industry/ professional sector. The updating of curriculum needs to be a constant feature in the field and the National Universities Commission (NUC) is the regulating body in charge of universities in Nigeria which oversees this function. Interaction between industries and universities is vital for the successful development of manpower. The key players in strengthening interaction between computer science/ IT institution and industry according to Ali (2008) are universities, industries and government (via the NUC).

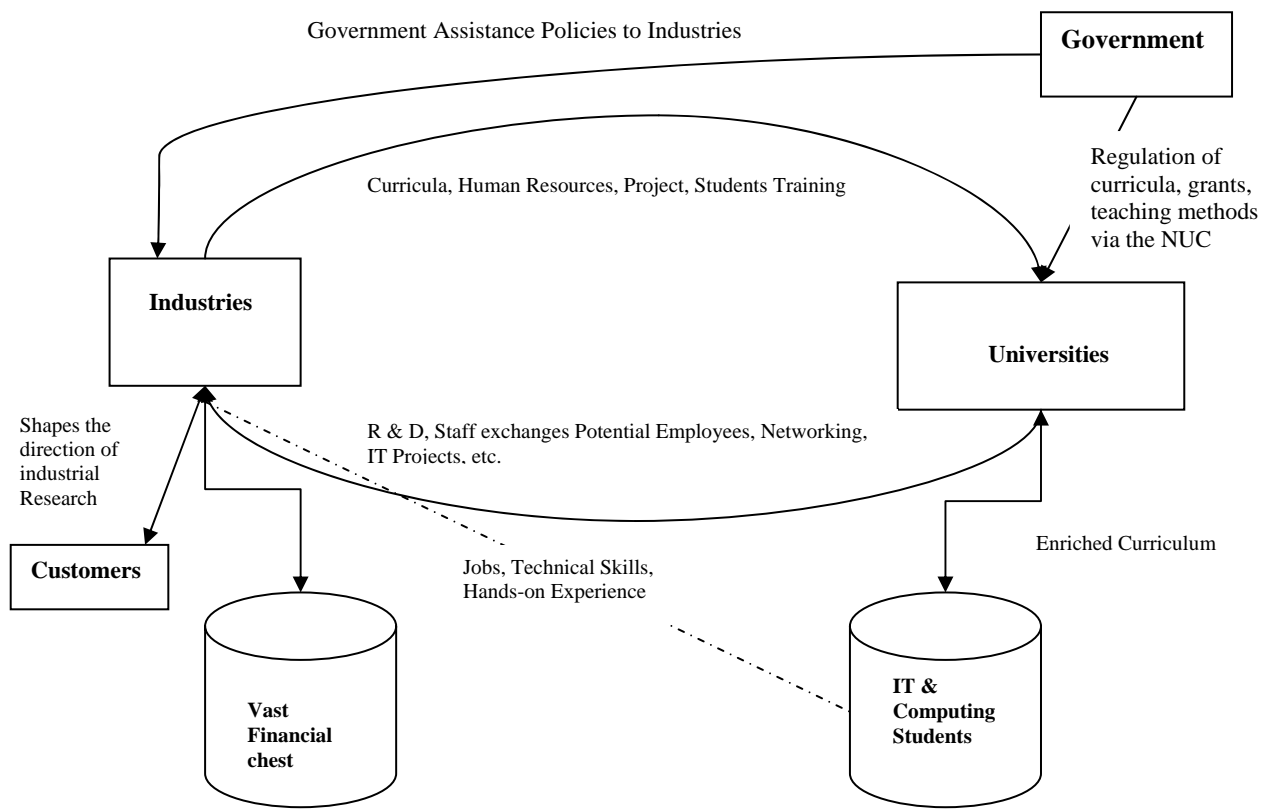

**Fig. 1: Suggested Model for Industry-University Partnership. Source.(Authors)**

In the above model (Fig. 1), the industries and partnering universities stand to gain a lot from each other and the relationship depicted is mutually-exclusive (beneficial to both partners). The industries, with vast access to finance through the shareholder funds, credits from financial institutions and operating capital, supplies the human resources or experts, get involved in curriculum development, initiates IT projects, pays for patents, etc. that will be used in exposing academic staff and students to the latest trends in IT and could even get involved in training academic staff and students while the partnering universities in turn, provide the potential employees who are tested and trained students, technical partners who are lecturers, completed IT projects which could be patented and commercialized to the industrial partner. Research and Development potentials and a chance to network with other universities and academic are also some of the benefit accruable to industries. The government in turn, regulates the universities through frequent accreditation of her curricula to meet with the dynamic nature of IT and gains too economically whenever more students are recruited for employment in the private sector through more manpower value being added to the nation's economy. It also provides assistance to industries through tax-rebates, conducive operating environments and other incentives that would encourage them to promote educational research and development. Finally, the customers determine the direction of industrial research since they provide the market base and revenue for industries.

## 3. BENEFITS OF IT-RELATED PARTNERSHIPS

In an IT-related partnership, the following are some of the benefits to both universities and industries alike:

| | Benefits for Industry | Benefits for Universities |
|---|---|---|
| a. | Thinking longer term by gaining an inside track on emerging trends and enabling technologies developed in universities | Improving market awareness by gaining insights into the research problems or interests to industries |
| b. | Benefiting from new ideas and past experience | Enriching teaching programs |
| c. | Going global by linking up the global academic networks | Maintaining research momentum in the IT sector |
| d. | Outsourcing through saving costs and letting universities handle research | Applying knowledge and skills to solving real business IT-related problems |
| e. | Access to IT skills within universities that company staff lack | Learning new IT skills and techniques developed in the industry |
| f. | Accessing a range of IT disciplines at once in a university | Learning new approaches to managing projects and how industry works |
| g. | Bringing additional financial resources to bear on research and thereby spreading costs | Drawing on a wider range of private funding and access to public funds requiring industry collaboration |
| h. | Reducing risk by sharing costs, finding out what others are doing | Building on excellence and reputation |
| i. | Complementing the company's physical resource base | Complementing the university's resource base |
| j. | Recruitment made easy | Sourcing job opportunities for IT graduates |

## CONCLUSION

In this paper, we have discussed the framework of an industry-university partnership in the area of information technology for Nigerian universities. The various types of partnerships and the overall benefits of such a partnership to both industries and universities alike were highlighted and the main goal of such a partnership was to help de-emphasise much dependence for funding on the government. The framework could be adapted to IT-related and non IT-related sectors and even to other countries in sub-Saharan Africa.